\documentclass[aps,prb,twocolumn,groupedaddress]{revtex4}

\usepackage{graphicx}
\usepackage{dcolumn}

\begin{document}


\title{Charge carrier correlation in the electron-doped $t$-$J$ model} 

\author{P. W. Leung}
\email[]{P.W.Leung@ust.hk}
\affiliation{Physics Dept., Hong Kong University of Science and Technology,
Clear Water Bay, Hong Kong}

\date{\today}

\begin{abstract}
We study the $t$-$t'$-$t''$-$J$ model with parameters chosen to model
an electron-doped high temperature superconductor.
The model with one, two and four charge carriers 
is solved on a 32-site
lattice using exact diagonalization.  
Our results demonstrate that at doping levels up to $x=0.125$ the model
possesses robust antiferromagnetic correlation. 
When doped with one charge carrier, the ground state has
momenta $(\pm\pi,0)$ and $(0,\pm\pi)$.
On further doping,
charge carriers are unbound and the momentum
distribution function can be constructed from that of the single-carrier ground
state. 
The Fermi surface resembles that of small pockets at
single charge carrier ground state momenta, which is the expected
result in a 
lightly doped antiferromagnet.
This feature persists upon doping up to the largest doping level we
achieved. 
We therefore do not observe the Fermi surface changing shape
at doping levels up to 0.125.
\end{abstract}


\maketitle

\section{Introduction}

It is well known that electron-doped high $T_c$ materials have very different
properties compared to hole-doped ones. 
Like hole-doped materials,
their undoped parent compounds are insulators with antiferromagnetic
spin order. 
But the electron-doped cuprate 
$\text{Nd}_{2-x}\text{Ce}_x\text{CuO}_4$ remains an
antiferromagnetic insulator up to doping level $x=0.13$ whereas in
the hole-doped cuprate $\text{La}_{2-x}\text{Sr}_x\text{CuO}_4$
a relative small doping level of $x\sim 0.02$ is enough to destroy its
antiferromagnetic correlation.\cite{d94} 
Theoretically it has been postulated that this asymmetry in
properties of electron- and hole-doped materials can be modeled by
adding intra-sublattice hopping terms to the $t$-$J$
model.\cite{tm94,gvl94} 
Compared to the nearest neighbor hopping motion in the $t$-$J$ model,
intra-sublattice hoppings
do not frustrate the spin background. Consequently the
antiferromagnetic order of the undoped system is better preserved upon
doping. 
Within this context,
various theoretical and numerical studies
have confirmed that the
electron-doped model has robust antiferromagnetic order. 
In addition, appropriate intra-sublattice hopping terms shift
single-carrier ground state momenta from $(\pm\pi/2,\pm\pi/2)$ in the
$t$-$J$ model to $(\pi,0)$ and its equivalent points.
This means that in a lightly doped system small charge carrier pockets will
form at $(\pi,0)$ and 
its equivalent points in
the first Brillouin zone. 
This agrees with angle-resolved
photoemission (ARPES) experiment\cite{a02,dhs03} on
$\text{Nd}_{2-x}\text{Ce}_x\text{CuO}_4$. 
Various properties of the electron-doped model\cite{tm01,tm03,t04,yclt04,ylt05}
including its
electronic states,
spin dynamics,
and Fermi surface evolution
have been worked out with emphasis on comparison
with experimental results.\cite{hubbard}

In this paper we are interested in the interaction among charge carriers
doped into the parent compound.
For this reason we conduct a systematic study on the
electron-doped model with a few charge
carriers using exact diagonalization. In this approach,
larger lattices are always preferred in order to minimize
finite-size effects. Furthermore, in order to study Fermi surface evolution
the lattice must have allowed {\bf k} points along the
antiferromagnetic Brillouin zone (AFBZ) boundary, i.e., from $(\pi,0)$ to
$(0,\pi)$. 
Square lattices having only 20 or 26 sites with periodic
boundary conditions do not have this property. 
The 32-site lattice is
the next available square lattice that has this property and on which
calculations are still manageable.
The $t$-$J$ model with up to four charge carriers on this lattice has
been studied in detail using exact diagonalization.\cite{lg95,clg98,l02,l05}
But previous calculations on the electron-doped model on this lattice 
have been limited to two
charge carriers only.\cite{ln23}
Since antiferromagnetic order in electron-doped cuprates is robust, we
need more charge carriers to make the antiferromagnetic phase unstable.
In this paper we report results for the electron-doped model with one,
two, and four 
charge carriers on a 32-site lattice, covering doping levels up to $x=0.125$.

Our paper is organized as follows.
We first define the model in section \ref{sec:he}. In section
\ref{sec:1e} we look at quasiparticle properties of a single charge
carrier doped into the system. Next we consider systems with multiple
charge carriers and
study the binding energies in section \ref{sec:be_e_dope} and the
real space charge carrier correlation in section \ref{sec:cr}. They provide the
first evidence that charge carriers are unbound at these doping levels.
Section \ref{sec:nq} deals with the momentum distributions of spin
objects which indicates the Fermi surface at different doping
levels.
In section \ref{sec:ss} we calculate the spin correlations which
clearly demonstrate that antiferromagnetic order persists upon doping,
and in section \ref{sec:jj} we attempt to search for other exotic
order when the antiferromagnetic correlation is weakened.
Finally we give our conclusion in section \ref{sec:conclusion}.

\section{Hole- and electron-doped models}
\label{sec:he}

We start from the extended $t$-$J$ model which was originally proposed to
describe hole-doped materials,
\begin{equation}
\label{tJ}
{\cal H}=-\sum_{\langle ij\rangle\sigma} t_{ij}(\tilde{c}^\dagger_{i\sigma}
\tilde{c}_{j\sigma}+\tilde{c}^\dagger_{j\sigma}\tilde{c}_{i\sigma})
+J\sum_{nn} \left({\bf S}_i\cdot{\bf S}_j
-\frac{1}{4}n_in_j\right).
\end{equation}
The nearest neighbor (nn)  spin exchange
constant $J$ is fixed  at $0.3$.
Farther than nearest neighbor hopping terms are necessary in order to
distinguish between hole and electron doping.
$t_{ij}=t,t',$ and $t''$ when $\langle ij\rangle$
is a pair of sites at distances 1, $\sqrt{2}$, and 2 apart
respectively, and is zero otherwise. 
The best fitting to ARPES results on
$\text{Sr}_2\text{CuO}_2\text{Cl}_2$ yields $t=1$, $t'=-0.3$, and
$t''=0.2$.\cite{lwg97}
In the case of hole doping,
$\tilde{c}^\dagger_{i\sigma}$ is a  spin (or electron) 
creation operator . To describe
electron-doped materials
it is usual to apply the
electron-hole transformation,\cite{gvl94}
$\tilde{c}_{i\sigma} \rightarrow \tilde{a}^\dagger_{i\sigma}$,
where $\tilde{a}^\dagger_{i\sigma}$ is a hole creation operator. The
resulting Hamiltonian for electron-doped materials is identical to
(\ref{tJ}) but with $\tilde{c}_{i\sigma}$ replaced by
$\tilde{a}_{i\sigma}$, etc, and $t_{ij}$ replaced by $-t_{ij}$. As a
result we can turn the Hamiltonian (\ref{tJ}) into 
an electron-doped model by flipping the signs of the hopping terms
$t_{ij}$.\cite{tm94}  
Despite this similarity, one should be reminded
that in the case of hole doping the operators $\tilde{c}_{i\sigma}$ in
the Hamiltonian are 
electron operators and the vacuum state is a state with no electron. The
condition of no double occupancy means that no more than one electron can
occupy the same site. At half-filling, each site has exactly one electron
and doping with holes means creating vacancies by removing
electrons. Translating 
into the language of the
electron-doped model, the operators $\tilde{c}_{i\sigma}$ are hole
operators and the vacuum 
state is a state with no hole, i.e., it cannot accommodate any more
electron. 
The condition of no double occupancy
means that each site can have no more than one hole. At half-filling,
each site has exactly one hole and doping means filling up
holes with electrons. 
To avoid confusion, we use the terms ``spin objects'' and ``charge
carriers'' to describe objects in the
Hamiltonian (\ref{tJ}).
In the case of hole doping, spin objects refer to electrons and charge
carriers refer to holes.
In the case of electron doping their meanings are reversed --- spin
objects refer to holes and charge 
carriers refer to electrons.
In this paper, by ``electron-doped model'' we
mean (\ref{tJ}) with hopping parameters $t=-1$, $t'=0.3$, and
$t''=-0.2$. [\onlinecite{edope_number}] 
In principle hole-doped model should refer to the same extended $t$-$J$ model
with $t=1$, $t'=-0.3$, and
$t''=0.2$.
But due to complications caused by
excited states of that model,\cite{l02} we choose to compare 
the electron-doped model with the simple $t$-$J$ model,
i.e., with $t=1$ and $t'=t''=0$. We remark that 
the  $t$-$J$ model was originally proposed to describe hole-doped
materials. But since they have different hopping terms, it will be
unfair to conduct a quantitative comparison between the $t$-$J$ and
electron-doped models. Instead we will mostly concentrate on their qualitative
differences. 

The electron-doped model with Hamiltonian (\ref{tJ}) and parameters
$J=0.3$, $t=-1$, $t'=0.3$, and
$t''=-0.2$ is
solved by exact diagonalization on a 
square lattice with 32 sites and periodic boundary conditions. 
Table \ref{tab:energies} shows the ground state energies and symmetries of the
electron-doped model with one, two, and
four electrons.
Calculations on the model with four electrons were performed on a cluster
of AMD Opteron servers with 64 CPUs.

 \begin{table}
 \caption{\label{tab:energies} Ground state energies, momenta and
   point group symmetries of the electron-doped
   model with $N_c$
   charge carriers.
$N_B$ is the number of basis in that particular
   subspace. The ground state energy at half-filling $E^0_0$ is $-11.329720$.}
 \begin{ruledtabular}
 \begin{tabular}{crdcc}
   $N_c$ &\multicolumn{1}{c}{$N_B$}
& \multicolumn{1}{c}{$E^{N_c}_0$} & {\bf k} & symmetry
      \\
\colrule
1 &   150,297,603 & -13.913616& $(\pi,0)$,$(0,\pi)$& $$\\
2 &   150,295,402 & -16.601689& $(0,0)$  & $d_{x^2-y^2}$\\
4 & 2,817,694,064 & -20.461647& $(0,0)$  & $s$\\
 \end{tabular}
 \end{ruledtabular}
 \end{table}

\section{Quasi-particle dispersion in the one-electron system}
\label{sec:1e}
\begin{figure*}
 \resizebox{16cm}{!}{\includegraphics{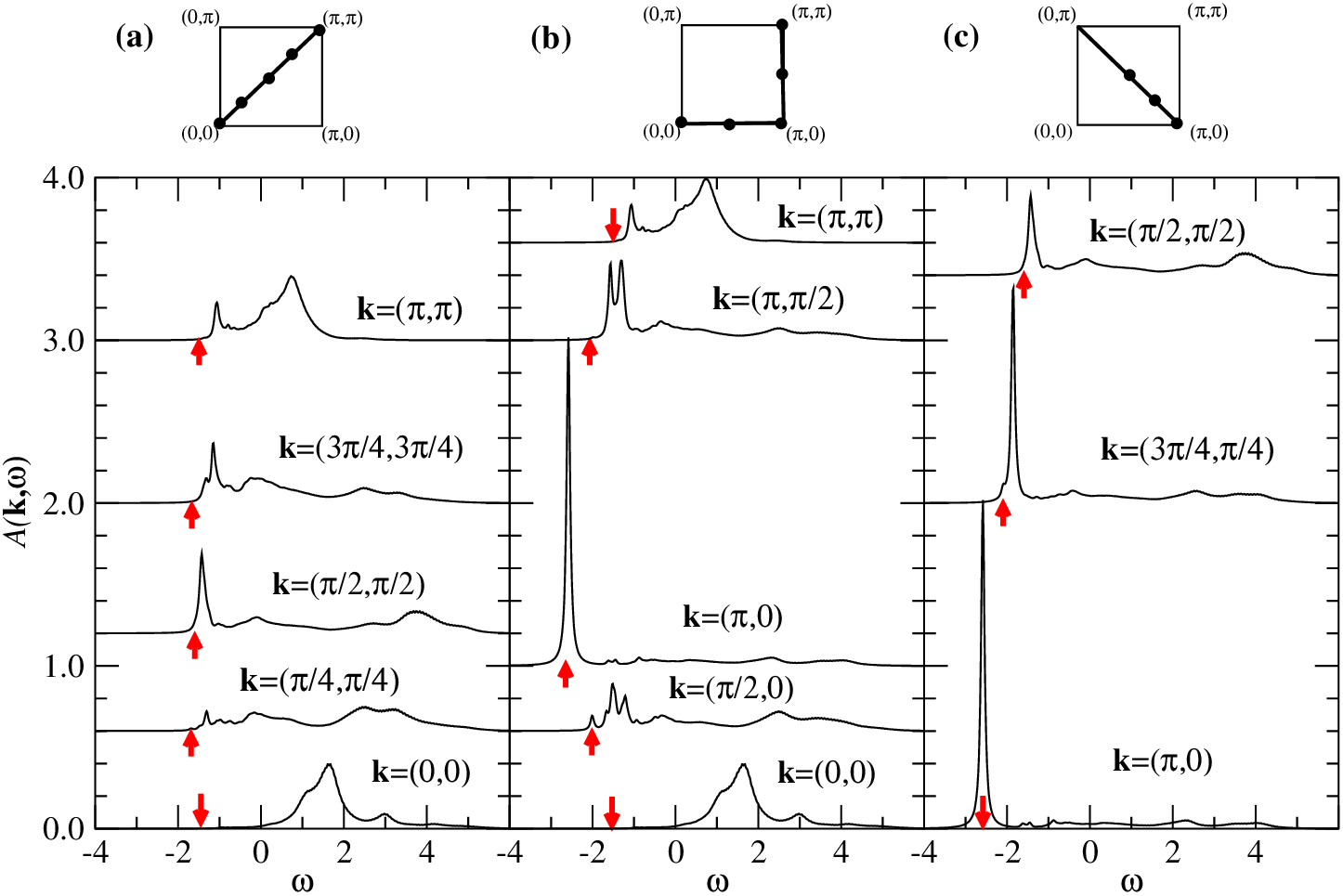}}
 \caption{\label{fig:aq} (Color online) Spectral function at
   half-filling $A({\bf 
     k},\omega)$ along three 
   branches: a) $(0,0)$ to 
   $(\pi,\pi)$, b) $(0,0)$ to $(\pi,0)$
   to $(\pi,\pi)$, and c) $(\pi,0)$ to $(0,\pi)$. Shaded arrows
   indicate the quasiparticle energy $E({\bf k})$.}
 \end{figure*}

\begin{figure}
 \resizebox{8cm}{!}{\includegraphics{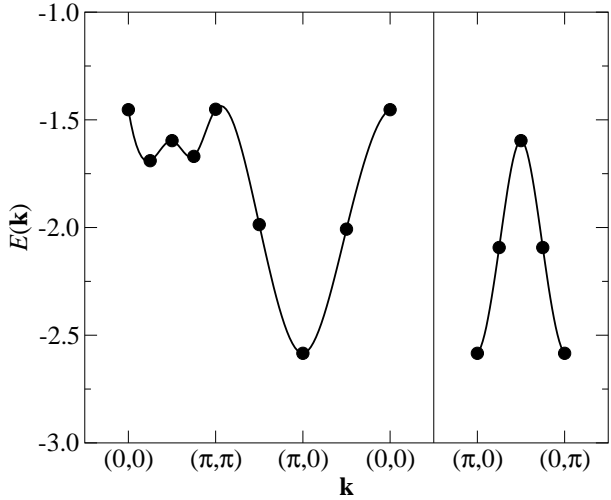}}
 \caption{\label{fig:dispersion} Quasiparticle dispersion
   relation obtained from the spectral function at half-filling in
     Fig.~\ref{fig:aq}. The line is a guide to the   
     eyes only.}
 \end{figure}

In the electron-doped model
the spectral function of spin objects at half-filling is defined as\cite{aq}
\begin{equation}
\label{aq}
A({\bf k},\omega)=\sum_n |\langle\psi^{1}_n|\tilde{c}_{{\bf
    k}\sigma}|\psi^0_0\rangle|^2 \delta(\omega -E^{1}_n+E^0_0),
\end{equation}
where $E^0_0$ and $\psi^0_0$ are the ground state energy and
wave function of the model at half-filling, and $E^{1}_n$ and
$\psi^{1}_n$ are energy and wave function of the $n$th excited state
of the model with one charge carrier. 
$A({\bf k},\omega)$ can be evaluated easily using the continued fraction
expansion. At each {\bf k} point we use 300 iterations and add an
artificial broadening factor $\epsilon=0.05$ to the delta function.
Fig.~\ref{fig:aq} shows $A({\bf k},\omega)$ along three branches in
the first Brillouin zone. At the Brillouin zone center
$(0,0)$, the spectral function has a broad structure.
It does not have any noticeable low energy peaks. 
As we move along the $(1,1)$ direction towards $(\pi,\pi)$
[Fig.~\ref{fig:aq}(a)] the 
spectral weight spreads to lower and higher energies, forming
well-defined peaks at low energies. When we go pass $(\pi/2,\pi/2)$,
the reverse occurs and at $(\pi,\pi)$, the spectral function has a
broad structure again. A similar trend is observed in the branch from
$(0,0)$ to $(\pi,\pi)$ through $(\pi,0)$ 
[Fig.~\ref{fig:aq}(b)]. This branch has the
largest dispersion. A different trend is observed along 
the AFBZ boundary, $(\pi,0)$ to $(0,\pi)$ [Fig.~\ref{fig:aq}(c)]. 
There the spectral weight mostly concentrates in
low energy states. As usual we define the quasiparticle weight $Z_{\bf
  k}$ as 
\begin{equation}
Z_{\bf k}=\frac{|\langle\psi^1_n|\tilde{c}_{{\bf
      k}\sigma}|\psi^0_0\rangle|^2}
{\langle\psi^0_0|\tilde{c}^\dagger_{{\bf k}\sigma}\tilde{c}_{{\bf
      k}\sigma}|
\psi^0_0\rangle},
\label{eq:zk}
\end{equation}
where $\psi^1_n$ is the lowest energy one-electron state which has
non-zero overlap with $\tilde{c}_{{\bf
    k}\sigma}|\psi^0_0\rangle$. These values are tabulated in
Table~\ref{tab:zq} together with the quasiparticle energy, 
\begin{equation}
E({\bf k})=E^1_n-E^0_0,
\end{equation}
which is the energy of the state $\psi^1_n$ in Eq.(\ref{eq:zk})
relative to the 
ground state energy at half-filling. 
Most $Z_{\bf k}$ are too
small to make their corresponding peaks visible in
Fig.~\ref{fig:aq}. Their positions are indicated
by shaded arrows. 
An obvious exception is  ${\bf k}=(\pi,0)$  which is
the ground state momentum of the one-electron system. There the
lowest energy peak has more than 60\% of the total
spectral weight. The quasiparticle energy is plotted in
Fig.~\ref{fig:dispersion}. The bandwidth is $3.777J$, which is in
qualitative 
agreement with the prediction of spin-polaron
calculation.\cite{bcs96} 

\begin{table}
\caption{\label{tab:zq} Quasiparticle energy and weight in the
   electron-doped model.}
\begin{ruledtabular}
 \begin{tabular}{cdd}
{\bf k}&\multicolumn{1}{c}{$E({\bf k})$}&\multicolumn{1}{c}{$Z_{\bf k}$}\\
\colrule
(0,0)                                  &-1.452767&0.00008  \\
($\frac{\pi}{4}$,$\frac{\pi}{4}$)      &-1.689698&0.00354  \\
($\frac{\pi}{2}$,$\frac{\pi}{2}$)      &-1.596649&0.00465  \\
($\frac{3\pi}{4}$,$\frac{3\pi}{4}$)    &-1.669586&0.00045  \\
($\pi$,$\pi$)                          &-1.450790&0.00005  \\
($\pi$,$\frac{\pi}{2}$)                &-1.986268&0.00237  \\
($\pi$,0)                              &-2.583895&0.63608  \\
($\frac{\pi}{2}$,0)                    &-2.007321&0.02755  \\
($\frac{3\pi}{4}$,$\frac{\pi}{4}$)     &-2.092922&0.01890  \\
\end{tabular}
\end{ruledtabular}
\end{table}

\section{Binding energy}
\label{sec:be_e_dope}
The $N_c$-charge carrier binding energy is defined as the excess energy of a
system with $N_c$ charge carriers over $N_c$ single-carrier systems,\cite{ry89}
\begin{equation}
\label{eqn:be1}
E_{N_cI}\equiv (E^{N_c}_0-E^0_0)-N_c(E^1_0-E^0_0).
\end{equation}
It indicates the tendency of the $N_c$ charge carriers to form a bound
state.  
From values in Table~\ref{tab:energies} we find that $E_{2I}=-0.1042$
and $E_{4I}=1.2037$. 
Compared to the $t$-$J$ model where $E_{2I}=-0.0515$,\cite{clg98}
it is tempting to interpret these numbers as evidence showing that
two charge carriers in the electron-doped model have
larger tendency to form a bound pair than in the $t$-$J$ model. 
However, we have two reasons to believe that this is not a fair
conclusion. First of all
we should not compare the tendency to form bound pairs in the two
models based on the magnitudes of their binding
energies because they have different hopping terms. As we shall see
in the next section, two charge carriers in the electron-doped model
in fact have a smaller tendency to form a bound state than in
the $t$-$J$ model.
Second, 
one must be careful in interpreting binding energies defined
in Eq.~(\ref{eqn:be1}) because they are very susceptible to
finite-size effects. Binding energies found in a finite system tend to
be lower than their true values in the thermodynamic limit. As already
pointed out in Ref.~\onlinecite{clg98}, we have no a priori reason to
believe that $E_{2I}$ can be extrapolated linearly in
$1/N$, where $N$ is the lattice size. 
Nevertheless, doing so with results at $N=16$ and 32 we obtain
$E_{2I}\sim -0.01$. This small value
already hinted that the charge carriers may not form a bound state.
When there are four charge carriers in the system, the large and
positive $E_{4I}$ clearly shows that they have no tendency to form a bound
state.

\section{charge carrier correlation in Real space}
\label{sec:cr}
\begin{figure}
 \resizebox{8cm}{!}{\includegraphics{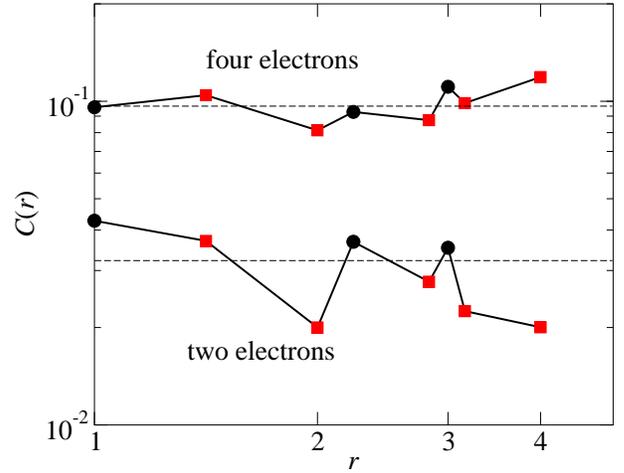}}
 \caption{\label{fig:cr} (Color online) Charge carrier correlation in systems
   doped with two and four electrons. Squares and
   circles indicate that the pair of electrons
   are in the same and opposite sublattices respectively. Dotted lines
   are pair correlations of uncorrelated electrons in
   the respective cases.}
 \end{figure}

The real space correlation among charge carriers can be clearly
displayed in the
charge carrier correlation 
function 
\begin{equation}
C(r)=\langle (1-n_r)(1-n_0)\rangle,
\end{equation}
where $n_r\equiv\tilde{c}^\dagger_r\tilde{c}_r$ is the number
operator of spin objects as in Eq.~(\ref{tJ}). 
Note that we use the convention\cite{clg98}
\begin{equation}
C(r)=\frac{1}{N_c N_E(r)}\sum_{ij}\bigl\langle (1-n_i)(1-n_j)
\delta_{|i-j|,r}\bigr\rangle,
\end{equation}
where $N_E(r)$ is the number of equivalent pairs with separation
$r$. In this convention the correlation function satisfies the
sum rule
\begin{equation}
\sum_{r>0}N_E(r)C(r)=N_c-1,
\end{equation}
and the probability for finding a pair of charge carriers at distance
$r$ apart is 
\begin{equation}
P(r)=N_E(r)C(r)/(N_c-1).
\end{equation}
Results in the two- and four-electron systems are shown in Fig.~\ref{fig:cr}.
Note that we use different symbols to distinguish between two groups
of correlations 
-- those between pairs
of electrons in the same and opposite sublattices. 
A striking feature in the two-electron system
is that the correlation between two electrons in
opposite 
sublattices does not decay significantly with $r$.
It is almost constant, implying that electrons in opposite
sublattices are mostly
uncorrelated. The correlation 
between two electrons in the same sublattice shows a very different trend.
It is comparatively smaller than that in the other group
and decays more significantly with
distance $r$. The overall probability of finding a pair of electrons
in the same and opposite sublattices are $0.3935$ and $0.6065$
respectively. From these results we conclude that in the two-electron
system, electrons
prefer to stay in opposite sublattices where they can move almost
independently of each other.
Obviously this 
results from the fact that intra-sublattice hopping terms $t'$ and
$t''$ in ${\cal H}$
do not frustrate the spin
background and therefore allow electrons to move more freely. 
When we increase the number of electrons to four, the
behaviors of the
two groups of correlation become very similar. They
show small fluctuations about the uncorrelated value $(N_c-1)/(N-1)$, which is
indicated by a dotted line in Fig.~\ref{fig:cr}. This shows that even
in the four-electron system the electrons are mostly uncorrelated.
The root-mean-square
separation between two electrons $\sqrt{\langle r^2\rangle}$ are
2.2786 and 2.4131 in the two- and four-electron systems
respectively. The
proximity of these values to the root-mean-square
separation between two uncorrelated electrons, 2.3827,
again suggests that electrons in our systems are almost uncorrelated.

\section{charge carrier correlation in momentum space}
\label{sec:nq}

Next we go to the momentum space and study the momentum
distribution function of spin 
objects $\langle n_{{\bf k}\sigma}\rangle
=\langle \tilde{c}^\dagger_{{\bf k}\sigma}\tilde{c}_{{\bf
k}\sigma}\rangle$.
One motivation for studying $\langle n_{{\bf k}\sigma}\rangle$ is to learn
about the Fermi surface of the model.
For this purpose it is important to realize that 
in $t$-$J$-like models the momentum distribution function has a
dome shape in the first Brillouin zone.
This feature results from
minimizing the kinetic energy\cite{ew93,eos94} and is not
related to the actual Fermi surface of the model. 
Nevertheless,
those {\bf k} points along the AFBZ boundary are not affected by this
kinematic effect. Therefore it is possible to extract information on
the Fermi surface from these {\bf k} points.
A second motivation for studying the momentum distribution function is
to see whether the single-carrier ground state is relevant to the
physics of the multiply-doped system. This has been a subject of
discussion in the  $t$-$J$ model.\cite{sh90,eos94,eo95}
Note that in the electron-doped model some authors choose to report the
momentum distribution function of charge carriers,
$\langle {\overline n}_{{\bf k}\sigma}\rangle 
=\langle \tilde{c}_{{\bf k}\sigma}\tilde{c}^\dagger_{{\bf
k}\sigma}\rangle$,
instead of $\langle n_{{\bf k}\sigma}\rangle$.
These distribution functions have the following properties:
\begin{eqnarray}
\langle n_{{\bf k}\sigma}\rangle + \langle {\overline n}_{{\bf
    k}\sigma}\rangle  
&=& n^{\text{max}}_{
    \sigma}\equiv(N_\sigma+N_c)/N,\label{bound}  \\
\sum_{\bf k}\langle n_{{\bf k}\sigma}\rangle &=& N_\sigma,\\
\sum_{\bf k}\langle {\overline n}_{{\bf k}\sigma}\rangle &=& N_c,
\end{eqnarray}
where $N_\sigma$ is the number of spin objects with spin $\sigma$.
In this section we will start with the single-electron momentum
distribution functions. We will show that they are qualitatively similar to
those in the $t$-$J$ model. Thus their features are
generic to $t$-$J$-like models.
We will then discuss momentum distribution functions of the two- and
four-electron systems. We will show that they can be constructed from
the single-carrier result and that
the Fermi surfaces are consistent
with small pockets at single-carrier ground state momenta.

\subsection{One-electron system}
\begin{figure}
   \resizebox{8cm}{!}{\includegraphics{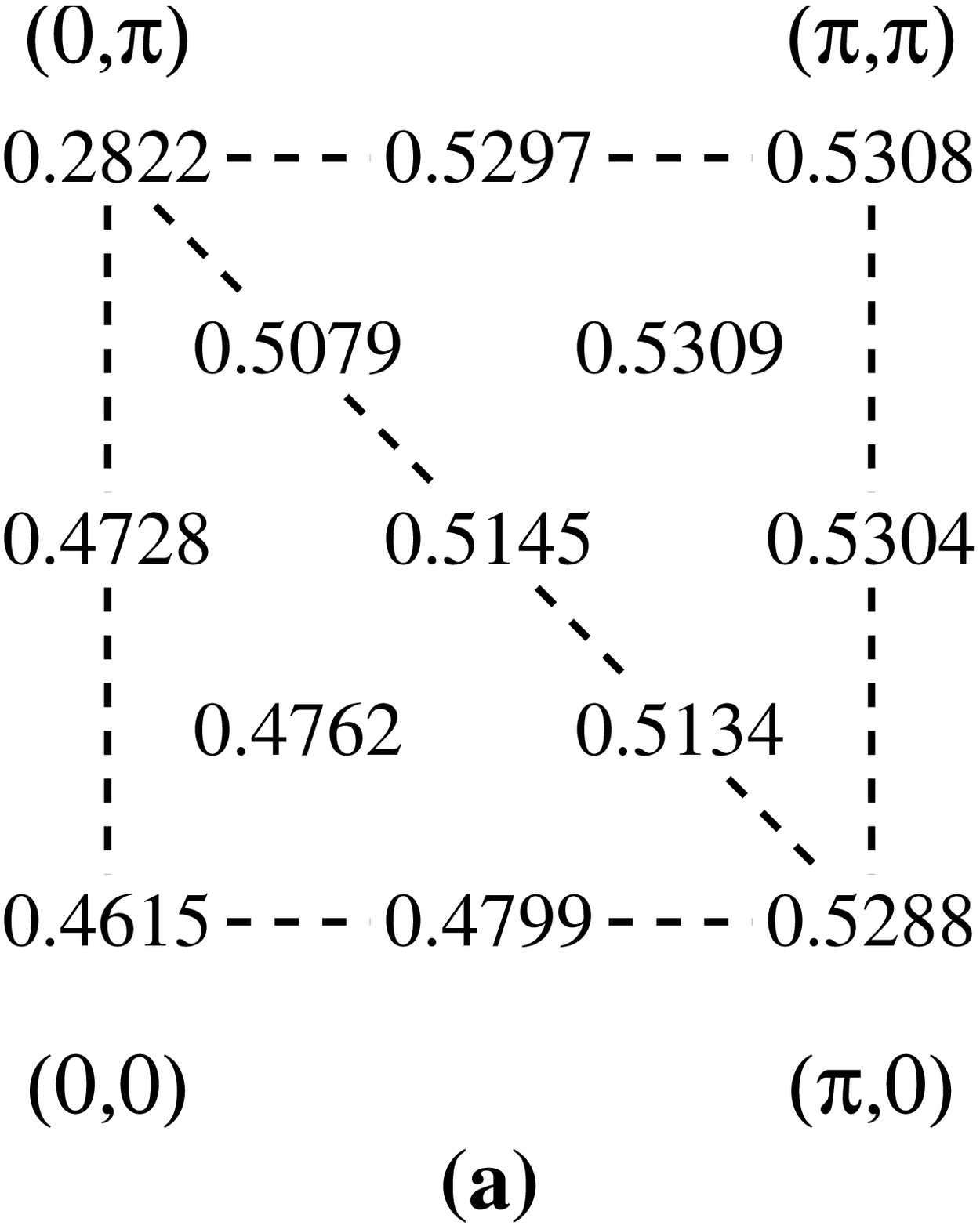}\hspace{1cm}
   \includegraphics{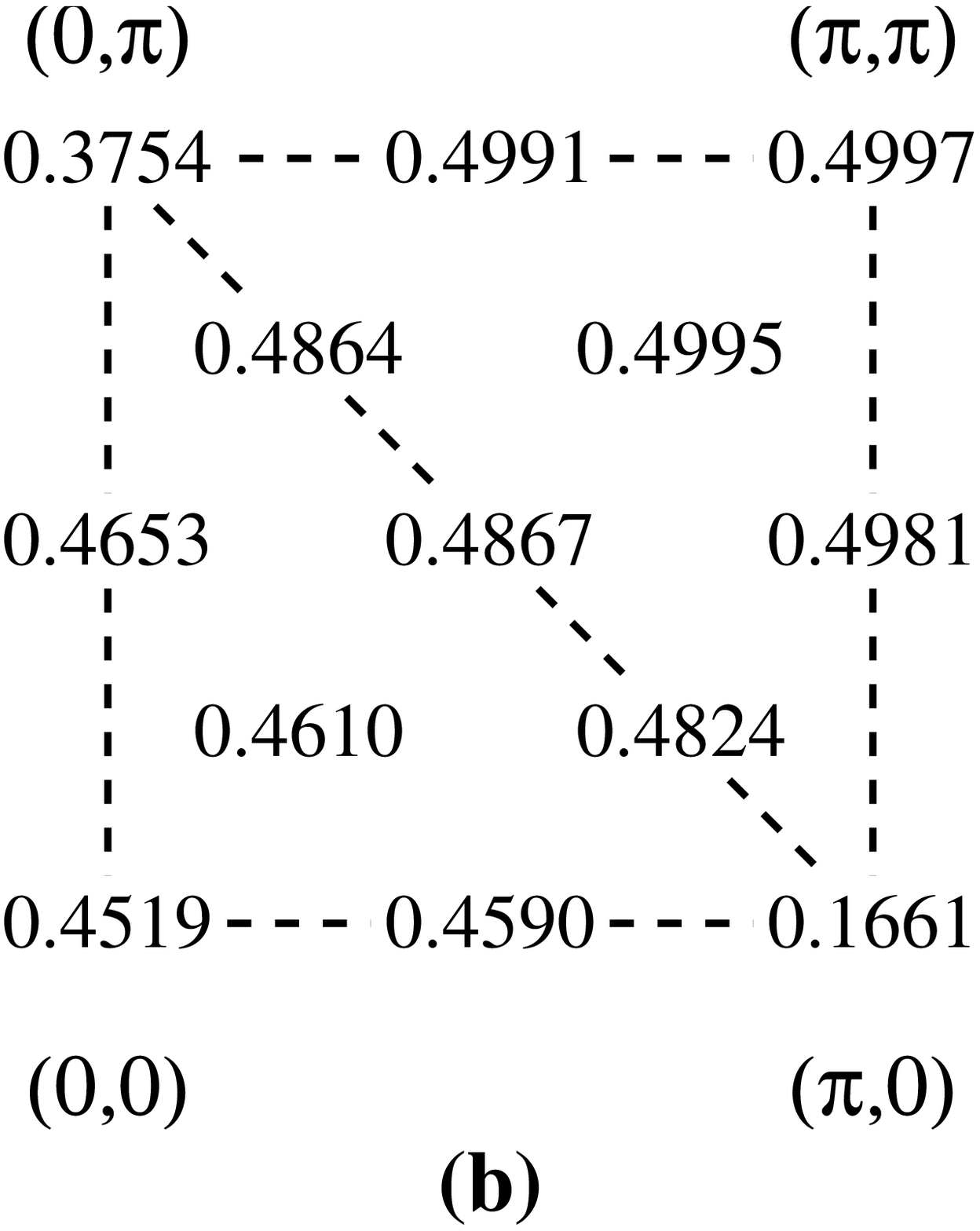}}  
 \caption{\label{fig:nq1e} Momentum distribution functions
   (a) $\langle n_{{\bf
   k}\uparrow}\rangle$ and (b) $\langle n_{{\bf
   k}\downarrow}\rangle$   in the ground state of the one-electron
   system with one spin-down object removed and  momentum $(\pi,0)$.
   Due to symmetry, only
   one quadrant of the Brillouin zone is
   shown. }
\end{figure}

Let us begin with the one-electron ground state with $S_z=1/2$ and
  momentum $(\pi,0)$. (Note that $S_z$ refers to the total spin of
  spin objects.)
The momentum distribution functions are
 shown in 
  Fig.~\ref{fig:nq1e}.
We immediately notice two very prominent features that exist in
both $\langle n_{{\bf k}\uparrow}\rangle$ and $\langle n_{{\bf
  k}\downarrow}\rangle$:
(i) there exist very sharp minima at $(\pi,0)$ or $(0,\pi)$;
(ii) besides these sharp minima, they are of a dome shape with a
  maximum around $(\pi,\pi)$ and slopes
  down towards a minimum at $(0,0)$.

As discussed above, the dome shape is a generic feature 
that also exists in the $t$-$J$ model.\cite{clg98}
The only difference
  is that the locations of the maximum and minimum of the
  ``dome'' are interchanged compared to those in the $t$-$J$ model. This is
  obviously due to the opposite signs of $t$ in the two models. 
This dome-shape feature
  therefore does not represent the shape of the true Fermi surface.
Note that $\langle n_{{\bf k}\uparrow}\rangle$ and
  $\langle n_{{\bf k}\downarrow}\rangle$ shift above and below the
  half-filled value of 1/2 respectively due to the
  restriction from Eq.~(\ref{bound}),
$\langle n_{{\bf k}\sigma}\rangle\leq n^{\text{max}}_{\sigma}$.

Similarly, sharp minima found in Fig.~\ref{fig:nq1e} are also found in
  the momentum distribution 
  functions 
of the $t$-$J$ model with one hole.
Just like in the $t$-$J$ model,
a  ``dip'' in $\langle n_{{\bf
  k}\downarrow}\rangle$ is found at the ground state
  momentum [$(\pi,0)$ in 
  Fig.~\ref{fig:nq1e}(b)], and an
  ``antidips'' in $\langle n_{{\bf
  k}\uparrow}\rangle$  is found at a {\bf k} point which is displaced from 
  the dip by the antiferromagnetic momentum
  $(\pi,\pi)$ [$(0,\pi)$ in
  Fig.~\ref{fig:nq1e}(a)]. 
From Fig.~\ref{fig:nq1e}(b) we find that the depth of the dip $\langle
  n_{{(\pi,\pi)}\downarrow}\rangle-\langle
  n_{{(\pi,0)}\downarrow}\rangle$ 
is 0.334. This is
  very close to $Z_{(\pi,0)}/2$ which is 0.318, indicating its close
  tie with the Fermi surface. Furthermore, $Z_{\bf k}$ along the edge
  of the AFBZ are very small. Therefore our data is consistent with a
  small Fermi surface, or carrier pocket, at $(\pi,0)$. This is to be
  expected in a lightly doped antiferromagnet.\cite{fulde}
Note that all features described so far are qualitatively the same in
  the hole- 
  and electron-doped models. 
They are generic features resulting from
  the kinematic effect and the antiferromagnetic order of the spin
  background. They do not
  reflect the different physics of 
  the hole- and electron-doped models.

\subsection{Systems with two and four electrons}

\begin{figure}
   \resizebox{8cm}{!}{\includegraphics{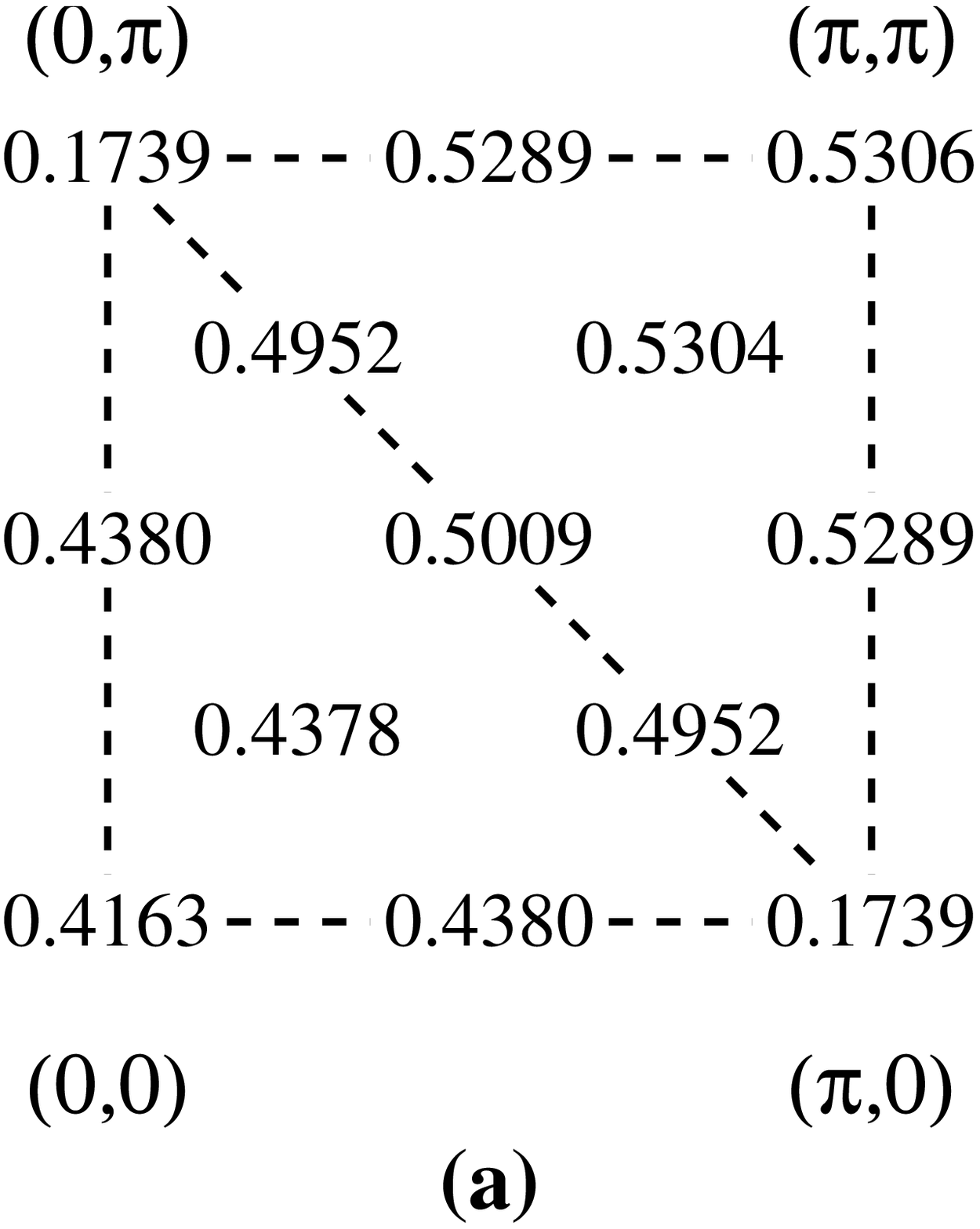}\hspace{1cm}
   \includegraphics{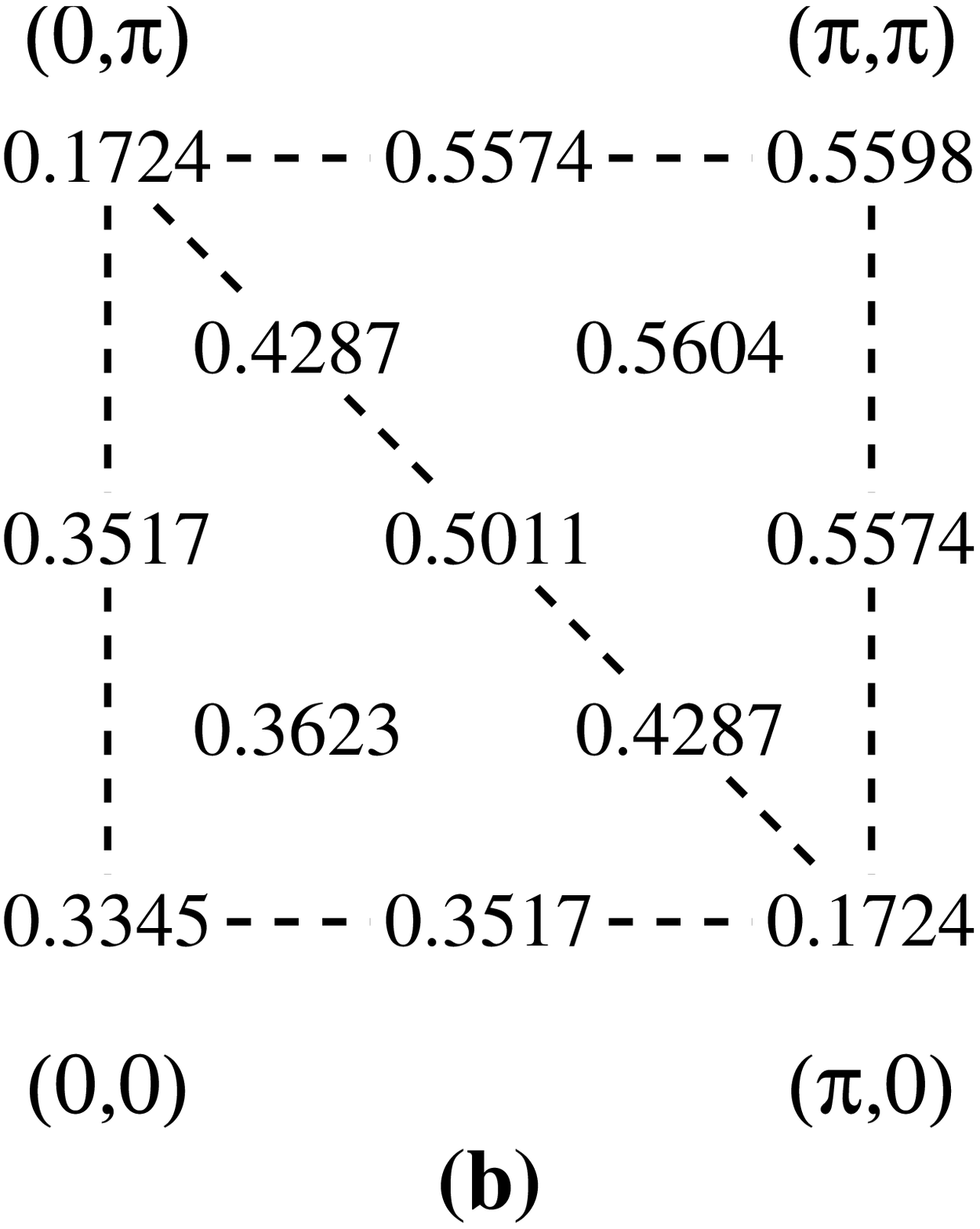}}  
 \caption{\label{fig:e_nq} Momentum distribution function $\langle n_{\bf
   k}\rangle$  in systems
   with (a) two and (b) four electrons. }
\end{figure}

When there is an even number of spin objects,
$\langle n_{{\bf
    k}\uparrow}\rangle=\langle n_{{\bf k}\downarrow}\rangle$ and we
drop the spin variable $\sigma$ from $\langle n_{{\bf
    k}\sigma}\rangle$. 
Fig.~\ref{fig:e_nq} shows $\langle n_{{\bf
    k}}\rangle$ in systems doped with two and four electrons. 
Again we can identity the same 
``dome-shape'' structure found in the one-electron system.
It is a generic feature of
the model and does not reflect the physics of the charge carriers. 
Further evidence for this comes from  the
``height'' of the dome, which is defined as $\Delta n\equiv \langle
n_{(\pi,\pi)}\rangle-\langle n_{(0,0)}\rangle$. In the two-electron
system  it is
0.114, which is roughly the same as $\Delta n_\uparrow + \Delta
n_\downarrow = 0.117$ in the one-electron system. And in the
four-electron system it is 0.225, roughly twice of that in the
two-electron system. 
These agreements show that the dome-shaped structures at different
doping levels are due to the same effect. 

Another feature common to the one-, two- and four-electron systems is
that their $\langle n_{\bf k}\rangle$ have very prominent dips at ${\bf
  k}=(\pi,0)$ and $(0,\pi)$. These dips resemble electron pockets.
This is certainly not a generic feature of $t$-$J$-like models because
it is not found in the $t$-$J$ model.\cite{clg98}
Note that $(\pi,0)$ and $(0,\pi)$ are
along the AFBZ boundary where $\langle n_{\bf k}\rangle$ is not
disguised by the 
generic dome-shape feature. Therefore these pocket-like features should
reflect the physics of the systems.
The fact that pocket-like features are found at these doping levels
immediately suggests the relevance of the singly-doped state to the
multiply-doped one. If electrons doped into the parent system behave
like weakly interacting fermions, then it is reasonable to expect that
multiple-electron systems can be approximated by filling up the
single-electron band. This should lead to electron-pockets at
single-electron ground state momenta.
To make this
argument quantitative, 
we consider the charge-carrier distribution function 
$\langle {\overline n}_{\bf
  k}\rangle$. From Eq.~(\ref{bound}), $\langle {\overline n}_{\bf
  k}\rangle$ can be considered as the suppression of $\langle {n}_{\bf
  k}\rangle$ from its maximum value upon doping. 
If multiple-electron states can be built up from single-electron
states, we expect the suppressions in $\langle {n}_{\bf  k}\rangle$
due to individual 
electrons doped into the system to be additive,
\begin{eqnarray}
\langle{\overline n}_{\bf k}\rangle_2 &\simeq& \langle{\overline n}_{{\bf
    k}\uparrow}\rangle_1 + \langle{\overline n}_{{\bf
    k}\downarrow}\rangle_1,\nonumber\\
\langle{\overline n}_{\bf k}\rangle_4 &\simeq& 2(\langle{\overline n}_{{\bf
    k}\uparrow}\rangle_1 + \langle{\overline n}_{{\bf
    k}\downarrow}\rangle_1,\label{eq:additive})
\end{eqnarray}
where $\langle{\overline n}_{\bf k}\rangle_{N_c}$ is the distribution
function of a system with $N_c$ electrons. 
Table~\ref{tab:nq} shows that in the two-electron system the additive
approximation  is
satisfied at all available {\bf k} points. 
In the four-electron system
it works satisfactorily at most {\bf k}
points. Obvious exceptions are 
$(\pi,0)$ and $(3\pi/4,\pi/4)$. 
Note that at this doping level
Eq.~(\ref{eq:additive}) cannot hold at $(\pi,0)$ without
violating Eq.~(\ref{bound}).
As a result,
instead of filling states at $(\pi,0)$  some electrons will
fill the next available low energy states which, according to
Fig.~\ref{fig:dispersion},  are at $(3\pi/4,\pi/4)$.
Therefore our results show that the additive approximation
works in
the electron-doped model at doping levels up to at least 0.125. Note
that this conclusion is consistent with that in
section~\ref{sec:be_e_dope}: if the electrons are uncorrelated, we
expect that multiply-doped states can be build up from the
singly-doped state.

\begin{table}
\caption{\label{tab:nq} Charge-carrier distribution function
   $\langle{\overline n}_{\bf k}\rangle$ in 
   systems doped with two and four electrons. Numbers in
   brackets are the corresponding results from the
   additive approximation
   Eq.~(\ref{eq:additive}).
   Note that in Eq.~(\ref{eq:additive})
   we have to average the one-carrier results over
   all degenerate ground states.}
\begin{ruledtabular}
 \begin{tabular}{ccc}
{\bf k}&
two-electron system&
four-electron system\\
\colrule
(0,0)                                  &0.1150 (0.1179) &0.2280 (0.2357)\\
($\frac{\pi}{4}$,$\frac{\pi}{4}$)      &0.0935 (0.0941) &0.2002 (0.1882)\\
($\frac{\pi}{2}$,$\frac{\pi}{2}$)      &0.0304 (0.0301) &0.0614 (0.0602)\\
($\frac{3\pi}{4}$,$\frac{3\pi}{4}$)    &0.0008 (0.0009) &0.0021 (0.0018)\\
($\pi$,$\pi$)                          &0.0006 (0.0008) &0.0027 (0.0015)\\
($\pi$,$\frac{\pi}{2}$)                &0.0024 (0.0021) &0.0051 (0.0042)\\
($\pi$,0)                              &0.3574 (0.3550) &0.3901 (0.7099)\\
($\frac{\pi}{2}$,0)                    &0.0932 (0.0928) &0.2108 (0.1855)\\
($\frac{3\pi}{4}$,$\frac{\pi}{4}$)     &0.0361 (0.0362) &0.1338 (0.0724)\\
\end{tabular}
\end{ruledtabular}
\end{table}

\section{Spin order}
\label{sec:ss}
\begin{figure}
   \resizebox{8cm}{!}{\includegraphics{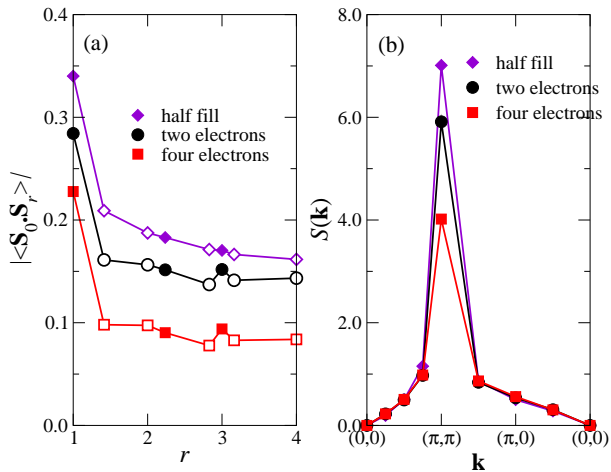}}  
 \caption{\label{fig:ss_sq} (Color online) (a) Spin correlation
   function and (b) 
   static structure 
   factor of systems with two
   and four electrons. Results at half-filling are given for reference
   purpose. In (a), empty and filled symbols represent
   positive and negative correlations respectively.}
\end{figure}

Our previous discussions have been based on the scenario that
antiferromagnetic spin order is preserved upon doping.
This is a direct
consequence of the fact that the $t'$ and
$t''$ hopping terms do not frustrate the spin background. In this
section we provide evidence for the existence of antiferromagnetic
correlation. 
Antiferromagnetic spin order can be measured directly using the spin
correlation function $\langle {\bf S}_0\cdot{\bf
  S}_r\rangle$. 
Results are shown in Fig.~\ref{fig:ss_sq}(a). 
At half-filling the system is known
to possess long-range antiferromagnetic order\cite{m91} and its spin
correlation 
is shown as a reference. We note that in the two-electron system the spin
correlation is not much weaker than that at half-filling, and more
importantly it does not show
significant decay beyond $r=\sqrt{2}$. This indicates that strong
antiferromagnetic spin order exists in the system.
The same qualitative trend is also observed in the four-electron system.
Although the spin correlation is inevitably weakened
due to higher doping level, it does not decay significantly beyond
$r=\sqrt{2}$. 
Another way to display the same data is through the static structure
factor,
\begin{equation}
S({\bf k})=\sum_{r} e^{i{\bf k}\cdot {\bf r}}\langle {\bf
  S}_0\cdot{\bf S}_r\rangle.
\end{equation}
Fig.~\ref{fig:ss_sq}(b) shows the structure factors in systems doped with
two and four  electrons. As
the doping level increases, the height of the antiferromagnetic
peak at $(\pi,\pi)$ is reduced. But it still remains prominent and
there is no sign of enhancement at any other {\bf k} point.
Our results therefore indicate that antiferromagnetic
order persists at least up to $x=0.125$. Note that this
doping level is close to the point where the antiferromagnetic phase
ends in the 
phase diagram of NdCeCuO,\cite{d94} which is $x=0.13$.

 \begin{figure*}
   \resizebox{14cm}{!}{\includegraphics{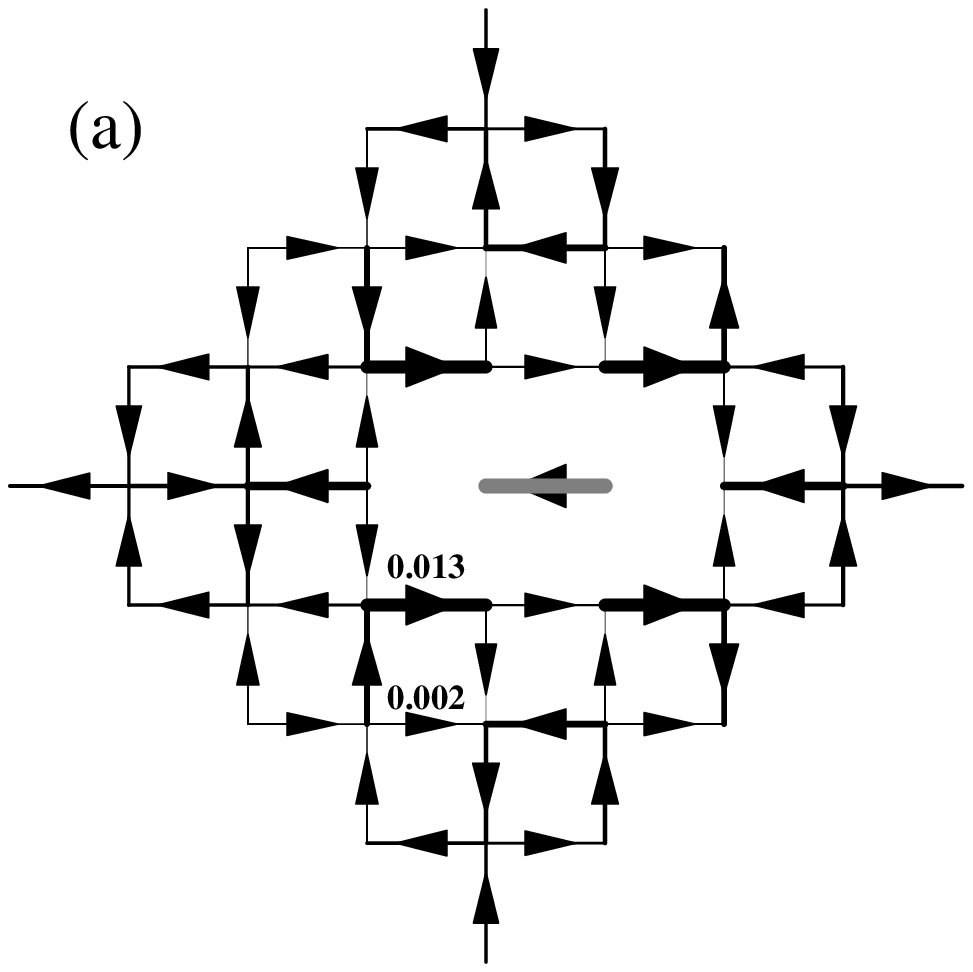}\hspace{1cm}
     \includegraphics{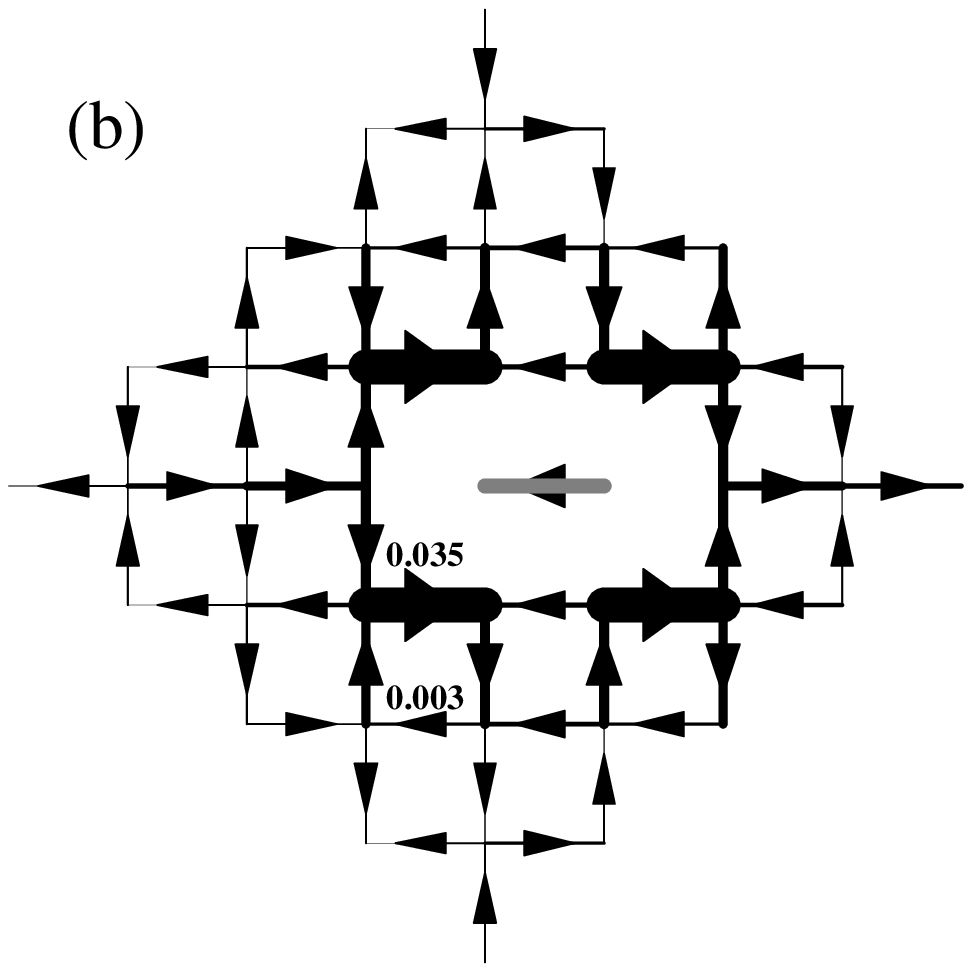}}   
  \caption{\label{fig:e-jj} Current correlation $\langle
   j_{kl}j_{mn}\rangle/x$
   in the ground state of the (a) two-electron,
   and (b) four-electron systems. The
   reference bond $mn$ is indicated by a shaded line. On other bonds,
   arrows point along the directions of positive correlation and
   line widths are proportional to $\langle
   j_{kl}j_{mn}\rangle/x$. For
   reference purpose, numerical values of $\langle
   j_{kl}j_{mn}\rangle/x$ 
   are shown next to some of the bonds.  }
\end{figure*}

 \begin{figure}
   \resizebox{8cm}{!}{\includegraphics{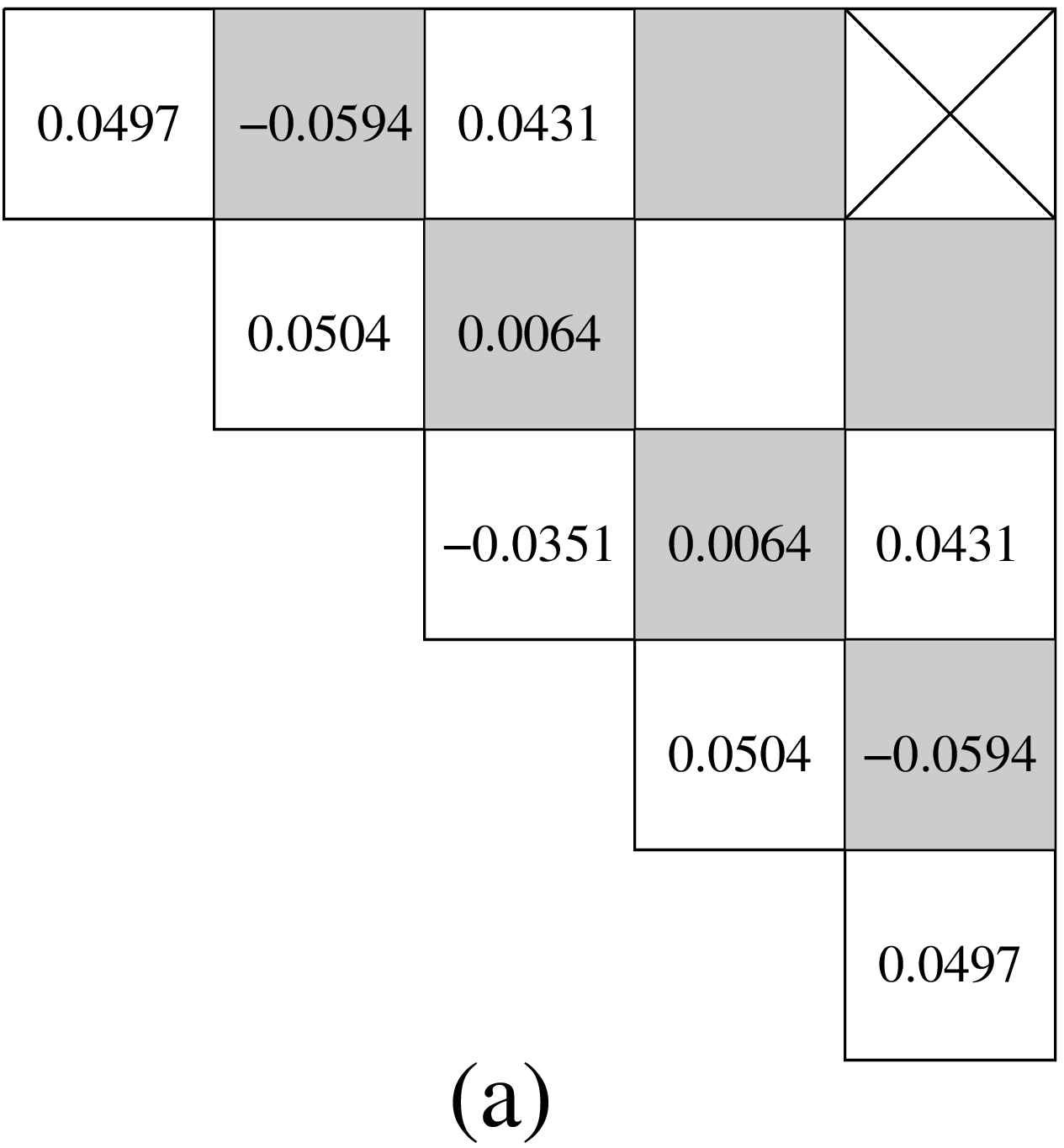}\hspace{0.5cm}
    \includegraphics{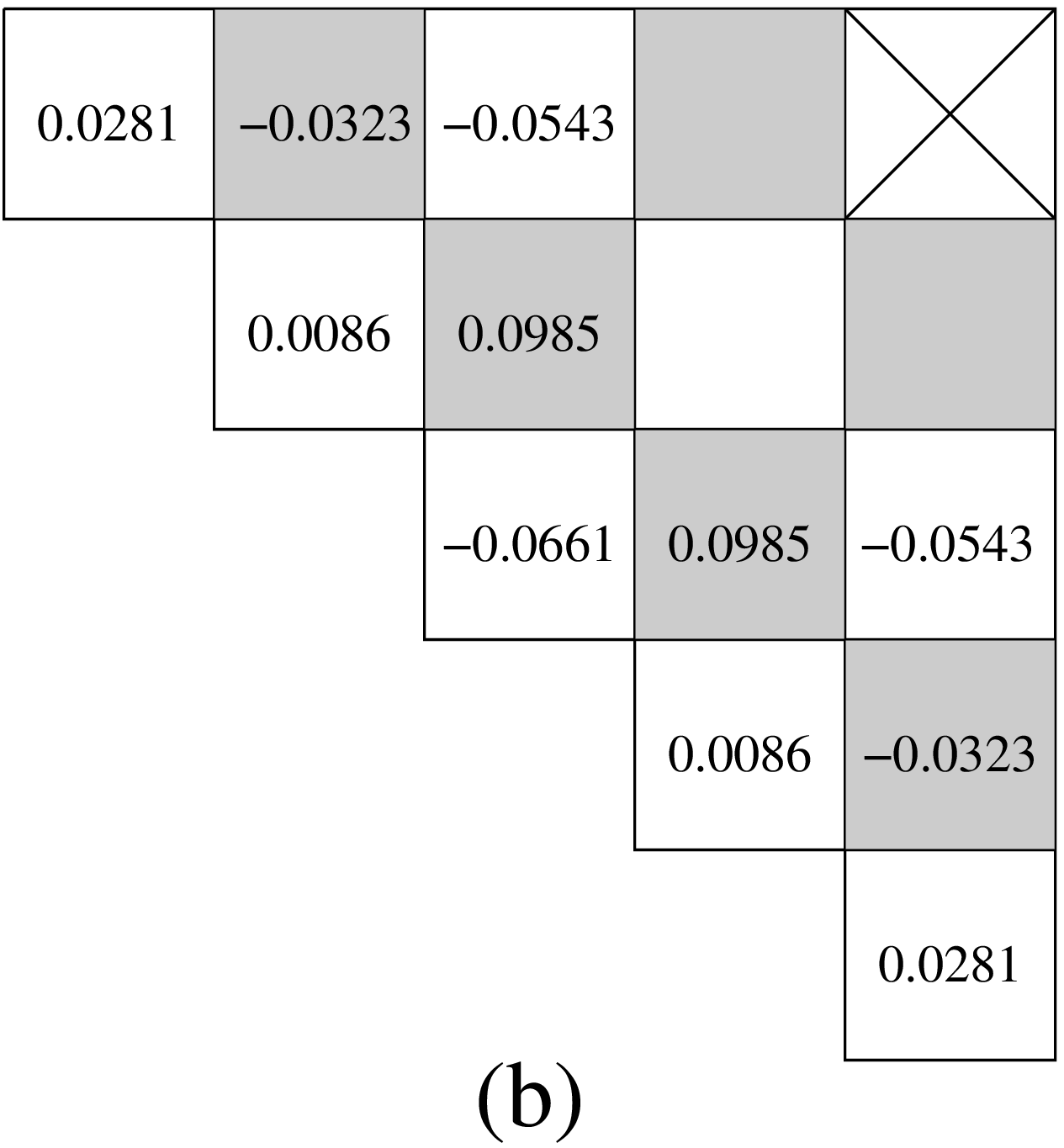}}   
  \caption{\label{fig:vorticity} Vorticity correlation $C_{VV}(r)$
     in the ground state of the (a) two-electron and (b) four-electron
     systems. The reference plaquette is indicated by a cross inside
     it. Those plaquettes touching the reference one do not have
     well-defined vorticity correlation because current operators on
     overlapping bonds do not commute.}
\end{figure}

\section{Charge current correlation}
\label{sec:jj}
The existence of a staggered pattern in the charge current correlation
function has been established in the $t$-$J$ model.\cite{ilw00,l00}
It has been interpreted as a direct evidence for
the staggered-flux phase in the mean-field picture.\cite{we01}
In this picture the hopping motions of charge carriers frustrate the
antiferromagnetic spin background and lead to staggered chiral spin
correlation. Binding of charge carriers is mediated through the
attraction between charge carriers with opposite
vorticity. Numerically it has been found that staggered current
pattern exists in the 
two-hole $t$-$J$ model with $d_{x^2-y^2}$ symmetry
only when the holes are loosely bound. It does not exist in states
with other symmetry, nor when $J/t$ 
is so large that the holes are tightly bound. The vorticity and charge
correlations are found to be proportional to each other.
However, in the electron-doped
model a very basic ingredient of the above picture is missing, namely,
the hopping motions of charge carriers are mostly unfrustrating. 
The result is that antiferromagnetic order is robust
and charge carriers do not have significant correlation. 
Consequently it is very unlikely that staggered current correlation
can exist. 
However, as antiferromagnetic spin order is
weakened at higher doping level, more subtle correlations may
emerge.\cite{rw03} We therefore calculate the current correlation
in the two- and four-carrier ground states and
see if there exists a systematic trend as doping level
increases.

Fig.~\ref{fig:e-jj} shows
the spatial variation of the current correlation function $\langle
j_{kl}j_{mn}\rangle$, where
\begin{equation}
j_{kl}=i(\tilde{c}^\dagger_k\tilde{c}_l-\tilde{c}^\dagger_l\tilde{c}_k),
\end{equation}
for systems doped with two and four electrons.
Another way to display the same data is to define the vorticity
$V({\bf r})$ 
of a square plaquette by summing up the current around it
in the counterclockwise direction.
The vorticity
correlation $C_{VV}(r)\equiv\langle V(r)V(0)\rangle/x$ is shown in
Fig.~\ref{fig:vorticity}. 
Our result for the two-electron system leaves little doubt that there
is no staggered
pattern in either $\langle j_{kl}j_{mn}\rangle$ or $C_{VV}(r)$.
In the four-electron system there is again no clear indication of
a staggered pattern.
We therefore conclude that intra-sublattice hopping terms in the
electron-doped model do not favor the formation of a staggered pattern
in the current 
correlation. This agrees with a recent mean-field study on the
electron-doped model.\cite{rw03} 
Furthermore, in the mean-field picture the loss of vorticity
correlation implies that charge carrier binding is not favored, which
 agrees with our results in section \ref{sec:cr}.
Finally we remark that in our four-electron system, 
it seems like at short distances the
current correlation is stronger and exhibit a staggered
pattern. This is most obvious when we compare
Fig.~\ref{fig:vorticity}(a) and (b). 
This seems to suggest that at doping
level $x=0.125$ the system may be starting to develop some other order
due to the weakening of antiferromagnetic correlation. 
However, the
range within which we observe the ``right'' correlation is too short
and the correlation is too weak
for us to decide whether it has any significance.
Therefore we are not able to make any definite statement concerning
this matter.

\section{Conclusion}
\label{sec:conclusion}
We have solved the electron-doped model with one, two, and
four charge carriers on a 32-site square lattice. 
Our results cover doping levels up to $x=0.125$.
In the electron-doped model, intra-sublattice hoppings of
charge carriers do not frustrate the spin background. 
Most of our results presented above can be understood as consequences of
this fact. 
Since hopping motions are mostly unfrustrating, charge carriers can
propagate more 
freely. This is reflected in the large
quasi-particle bandwidth in the singly-doped system. 
In systems doped with two and four electrons, the charge carrier
correlation function shows that electrons are
uncorrelated. Again this is due to the fact that electrons can hop
more freely in the same sublattice. There is no evidence of charge
carriers forming a bound state.
Unfrustrating hopping motions also mean that antiferromagnetic
correlation in the spin background is better preserved upon doping. This
is clearly shown in the spin correlation function and static structure
factor. It also shows up in the Fermi surface of the system.
In the singly-doped system quasi-particle weights at {\bf k} points
along the AFBZ boundary are small except
at the single-carrier ground state momenta, i.e., $(\pi,0)$ and its
equivalent 
points. This resembles a Fermi surface consisting of small pockets at
single-carrier ground state momenta, which is expected in a
lightly doped antiferromagnet.\cite{fulde}
These small pockets persist in our systems doped with two and four
electrons and are clearly visible in their momentum distribution
functions of spin 
objects. Furthermore, momentum distribution
functions of our multiply-doped systems
can be well approximated by adding up the
singly-doped momentum distribution functions. This re-assures that
charge carriers are uncorrelated.

Our results show that antiferromagnetic order in electron-doped model
persists at least up to doping level $x=0.125$. We find no clear
evidence of 
other orders existing in our systems.
This is consistent with the
phase diagram of $\text{Nd}_{2-x}\text{Ce}_x\text{CuO}_4$ whose
antiferromagnetic phase persists up to $x=0.13$.\cite{d94}  
ARPES experiment on the same material shows that before the Fermi
surface becomes a large one centered at $(\pi,\pi)$, small pockets
will start to appear at $(\pi/2,\pi/2)$ (and its equivalent points)  
as those at $(\pi,0)$
evolve.\cite{dhs03} 
However, the quasi-particle energy $E({\bf k})$ at ${\bf
  k}=(\pi/2,\pi/2)$ in the electron-doped model is quite high (see
Fig.~\ref{fig:dispersion}). 
Assuming that the quasi-particle dispersion relation does not change
much on further doping,
electrons doped into the
system will fill lower energy states at other {\bf k} points first.
Therefore it is not surprising that we do not see pockets developing at
$(\pi/2,\pi/2)$.
A recent theoretical calculation predicts that pockets will start to
appear at $(\pi/2,\pi/2)$ at $x=0.144$ when the Fermi level crosses a
different band from that at lower doping levels.\cite{yclt04} 
Therefore it is possible that our doping level is
not large enough to observe the change in the Fermi surface as
revealed by ARPES experiment.

\begin{acknowledgments}
This work was supported by a grant from the Hong Kong Research Grants
Council (Project No. HKUST6159/01P).
\end{acknowledgments}


\begin{thebibliography}{}

\bibitem{d94}
E. Dagotto, Rev. Mod. Phys. {\bf 66}, 763 (1994).

\bibitem{tm94}
T. Tohyama and S. Maekawa, Phys. Rev. B {\bf 49}, 3596 (1994).

\bibitem{gvl94}
R. J. Gooding, K. J. E. Vos, and P. W. Leung, Phys. Rev. B {\bf 50},
12866 (1994).

\bibitem{a02}
N. P. Armitage, F. Ronning, D. H. Lu, C. Kim, A. Damascelli,
K. M. Shen, D. L. Feng, H. Eisaki, Z.-X. Shen, P. K. Mang, N. Kaneko,
M. Greven, Y. Onose, Y. Taguchi, and Y. Tokura, Phys. Rev. Lett. {\bf
  88}, 257001 (2002).

\bibitem{dhs03}
A. Damascelli, Z. Hussain, and Z.-X. Shen, Rev. Mod. Phys. {\bf 75},
473 (2003).

\bibitem{tm01}
T. Tohyama and S. Maekawa, Phys. Rev. B {\bf 64}, 212505 (2001).

\bibitem{tm03}
T. Tohyama and S. Maekawa, Phys. Rev. B {\bf 67}, 92509 (2003).

\bibitem{t04}
T. Tohyama, Phys. Rev. B {\bf 70}, 174517 (2004).

\bibitem{yclt04}
Q. Yuan, Y. Chen, T. K. Lee, and C. S. Ting, Phys. Rev. B {\bf 69},
214523 (2004).

\bibitem{ylt05}
Q. Yuan, T. K. Lee, and C. S. Ting, Phys. Rev. B {\bf 71}, 134522
(2005). 

\bibitem{hubbard}
We remark that there is another approach to the electron-doped model
which, in contrast to the strong coupling limit assumed in the $t$-$J$
model, emphasizes the importance of intermediate coupling.  See, for example,
D. S{\'e}n{\'e}chal and A.-M. S. Tremblay, Phys. Rev. Lett. {\bf 92},
126401 (2004), and references there-in.

\bibitem{lg95}
P. W. Leung and R. J. Gooding, Phys. Rev. B {\bf 52}, R15711 (1995).

\bibitem{clg98} 
A. L. Chernyshev, P. W. Leung, and R. J. Gooding, Phys. Rev. B {\bf
  58}, 13594 (1998).

\bibitem{l02}
P. W. Leung, Phys. Rev. B {\bf 65}, 205101 (2002).

\bibitem{l05}
P. W. Leung, cond-mat/0511568.

\bibitem{ln23}
P. W. Leung and K. K. Ng, Int. J. Mod. Phys. B {\bf 17}, 3367 (2003).

\bibitem{lwg97}
P. W. Leung, B. O. Wells, and R. J. Gooding, Phys. Rev. B {\bf 56},
6320 (1997).

\bibitem{edope_number}
It has been suggested that in the electron-doped model not
only the signs and the magnitudes of the hopping terms are
different from those in the hole-doped model (see
Ref.~\onlinecite{eskes}). But since we are not interested in the detail
quantitative differences, we only change the signs of the
hopping terms in the electron-doped model. 

\bibitem{aq}
Note that in the electron-doped model this definition of $A({\bf
  k},\omega)$ corresponds to inverse photoemission spectroscopy on the
undoped insulator. A similar definition in the $t$-$J$ model
corresponds to photoemission spectroscopy, see Ref.~\onlinecite{lwg97}.

\bibitem{bcs96}
V. I. Belinicher, A. L. Chernyshev, and V. A. Shubin, Phys. Rev. B
{\bf 53}, 335 (1996).

\bibitem{ry89}
J. A. Riera and A. P. Young, Phys. Rev. B {\bf 39}, R9697 (1989).

\bibitem{ew93}
R. Eder and P. Wr\'{o}bel, Phys. Rev. B {\bf 47}, 6010 (1993).

\bibitem{eos94}
R. Eder, Y. Ohta, T. Shimozato, Phys. Rev. B {\bf 50}, 3350 (1994).

\bibitem{sh90}
W. Stephan and
P. Horsch, Phys. Rev. Lett. {\bf 66}, 2258 (1990).

\bibitem{eo95}
R. Eder
and Y. Ohta, Phys. Rev. B {\bf 51}, 6041 (1995).

\bibitem{fulde}
P. Fulde, \textit{Electron Correlations in Molecules and Solids},
$3^{rd}$ edition, (Springer, Berlin, 1995).

\bibitem{m91}
E. Manousakis, Rev. Mod. Phys. {\bf 63}, 1 (1991).

\bibitem{ilw00}
D. A. Ivanov, P. A. Lee, and X.-G. Wen, Phys. Rev. Lett. {\bf 84},
3958 (2000). 

\bibitem{l00}
P. W. Leung, Phys. Rev. B {\bf 62}, R6112 (2000).

\bibitem{we01}
We remark that there is another independent interpretation of the
same result in 
the spin-polaron picture. See
P. Wr\'{o}bel and R. Eder, Phys. Rev. B {\bf 64}, 184504 (2001).

\bibitem{rw03}
T. C. Ribeiro and X.-G. Wen, Phys. Rev. B {\bf 68}, 24501 (2003).

\bibitem{eskes}
H. Eskes and G. A. Sawatzky, Physica C {\bf 160}, 424, (1989).

\end{thebibliography}
\end{document}